\documentclass[a4paper]{article}
\usepackage{graphicx}

\newcommand{\cA}{{\cal A}}

\newcommand{\cR}{{\cal R}}

\newcommand{\be}{\begin{equation}}
\newcommand{\ee}{\end{equation}}
\newcommand{\bea}{\begin{eqnarray}}
\newcommand{\eea}{\end{eqnarray}}

\newcommand{\f}{w}

\begin{document}
\vspace{-1.5cm}
\title{
   \begin{flushright} \begin{small}
     LAPTH-901/2002  \\ DTP-MSU/02-04 \\
  \end{small} \end{flushright}
{\bf
Monopoles in NBI-Higgs theory and Born-Infeld collapse}%
\author{ V.V. Dyadichev$^{a}$\thanks{Email: vlad@grg1.phys.msu.su}
      and  D.V. Gal'tsov$^{a,b}$\thanks{Email: galtsov@grg.phys.msu.su} \\ \\
$^{a}$Department of Theoretical Physics,\\
     Moscow State University, 119899, Moscow, Russia,\\
$^{b}$Laboratoire de  Physique Th\'eorique LAPTH (CNRS), \\
B.P.110, F-74941 Annecy-le-Vieux cedex, France}
}
\date{\today}
\maketitle
\begin{abstract}
Regular magnetic monopoles in the non-Abelian Born-Infeld-Higgs
theory are known to exist in the region of the field strength
parameter $\beta>\beta_{{\rm cr}}$, bounded from below. Beyond
this region, only pointlike (embedded abelian) monopoles exist,
and we show that the transition from the regular to singular
structure is reminiscent of gravitational collapse. Near the
threshold behavior is characterized by the rapidly increasing
negative pressure, which typically arises in the high density NBI
matter. Another feature, shared both the NBI and gravitating
monopoles, is the existence of excited states, which can be
thought of as bound states of monopoles and sphalerons. These are
labeled by the number $N$ of nodes of the Yang-Mills function.
Their masses are greater than the mass of the ground state
monopole, and they are expected to be unstable. The sequence of
masses $M_N$ rapidly converges to the mass of the embedded
Abelian solution with constant Higgs. The ratio of the sphaleron
size to that of the monopole grows with decreasing $\beta$, and,
at the same time, both fall down until the solutions cease to
exist, again exhibiting collapse to the pointlike monopole. The
results are presented and compared both for the ordinary and the
symmetrized trace NBI actions.
\end{abstract}

\newpage
\section{Introduction}
An effective low-energy dynamics of strings attached to multiple
D-branes is governed by the non-Abelian Born-Infeld (NBI)
action~\cite{Ts97,GiKu98}. Different aspects of this theory were
studied recently including the problem of magnetic monopoles. The
NBI theory contains a parameter, $\beta$, of dimension of the
field strength (string tension), reflecting a non-local nature of
the underlying string theory. It was observed
numerically~\cite{GrMoSc99,Gretal99} that monopole solutions
exist for $\beta$ varying from infinity (which corresponds to the
usual form of the YM action) till some boundary value $\beta_{\rm
cr}$, while for $\beta <\beta_{\rm cr}$ only pointlike (embedded
abelian) monopoles exist (see also \cite{Tr99,Tr99a}). Previous
study~\cite{GaDy00} also revealed an existence of the excited
monopoles in the NBI theory with the similar threshold behavior.
Here we study monopoles near the threshold in more details, and
find an interesting new phenomenon in the NBI theory, resembling
gravitational collapse.

In fact, magnetic monopoles in the Yang-Mills-Higgs (YMH) theory with
the standard YM lagrangian coupled
to gravity exhibit similar features~\cite{BrFoMa95}
(for a review and further
references see~\cite{VoGa98,Ga01}).  Gravity imposes an upper limit on the
monopole mass, whose physical meaning is simple: when the monopole radius
becomes smaller than its gravitational radius,
the black hole should be formed.
A detailed study of the monopole --- black hole transition was undertaken
recently by Lue and Weinberg~\cite{LuWe99,LuWe00}. The ratio of the
monopole mass to the Planck mass serves as an order parameter, $\alpha$,
describing  transition from the globally regular solutions to black holes.
Regular solutions exist for $\alpha<\alpha_{\rm cr}$, the critical value
$\alpha_{\rm cr}$ being of the order unity.  Another feature of
the near critical region is the bifurcation of the mass curve, giving rise
to new branches of the gravitationally excited monopoles. The latter may be
regarded as bound states of monopoles with gravitational
sphalerons~\cite{GaVo91}, known as Bartnik-McKinnon's (BK)
solutions~\cite{BaMc88}, which live inside the monopole core. Therefore, the
threshold in the parameter space of gravitating monopoles is associated both
with the monopole-black hole transition and branching off the gravitationally
excited monopoles.

It turns out that the flat space NBI theory admits sphaleron solutions
similar to Bartnik-McKinnon particles~\cite{GaKe99} (though with some
differences in the detailed structure~\cite{DyGa00,DyGa'00}). Moreover, when
the NBI theory is coupled to gravity, parameters of the NBI sphalerons are
continuously driven with the increasing Newton constant to these of the BK
particles. Existence of both the BK and the NBI sphalerons is related to
breaking of scale invariance of the Yang-Mills theory by non-linearity of the
BI lagrangian in the first case, and by gravity in the second. Therefore,
the possibility of bound states of the NBI sphalerons with magnetic monopoles
could be expected by an analogy with the gravitating case. Note that the
limit of vanishing gravity corresponds to the YM limit $\beta\to \infty$
in the case of the NBI theory.

Surprisingly enough, the analogy with gravitation goes farther, and one
observes some analog of gravitational collapse in the
flat space NBI theory. The size of monopoles (including excited states)
is rapidly decreasing when $\beta\to\beta_{\rm cr}$ with the limiting
configuration being an Abelian pointlike monopole. An analysis of stresses
inside monopoles near the criticality shows that both  radial and
tangential pressure become negative in the core and rapidly
increase. It is expected that in the dynamical picture these
stresses will force the regular configuration to shrink to the pointlike
structure.

The plan of the paper in as follows. In Sec.~2 we discuss the form of the NBI
lagrangian and present the ordinary trace and the symmetrized trace versions of
the theory. Sec.~3 contains the description of the monopole-sphaleron bound
states in the ordinary trace model. In Sec.~4 we discuss the structure of
solutions near the threshold and reveal the collapse behavior for both the
ground and excited monopoles. In sec.~5 the analogous results
are briefly discussed within the symmetrized trace NBI model.

\section{The action}
A subtle point in the definition of the NBI action is the specification of the
trace over the gauge group generators
\cite{Ts97,GaGoTo98,Br98,BrPe98,Pa99,Za00} (for an earlier discussion see
\cite{Ha81}). Formally a number of possibilities can be envisaged. Starting
with the determinant form of the $U(1)$ Dirac-Born-Infeld action
\begin{equation}  \label{det}
S=\frac{1}{4\pi}\int
\left\{1-\sqrt{-\det(g_{\mu\nu}+F_{\mu\nu})}\right\}d^4x,
\end{equation}
one has different options including the usual trace,
the symmetrized  or antisymmetrized traces \cite{Ts97},
or   evaluation of the determinant both with
respect to Lorenz and the gauge matrix indices \cite{Pa99}.
Alternatively, one can start with the Abelian `square root' form,
which is obtained in the four-dimensional case evaluating the
determinant under the square root:
\begin{equation}\label{squ}
\sqrt{-\det(g_{\mu\nu}+F_{\mu\nu})}=
\sqrt{-\det(g)}\;\sqrt{1+\frac12 F^2 -\frac{1}{16}(F\tilde F)^2},
\end{equation}
with $F^2=F_{\mu\nu} F^{\mu\nu},\;
F\tilde{F}=F_{\mu\nu}\tilde{F}^{\mu\nu}$. For a non-Abelian gauge
group this relation is no more valid, unless a special trace
prescription is made, but this U(1) action may serve as another
starting point for non-Abelian generalization applying the simple trace
to the right hand side of this equation.

A particular trace operation, for which the relation Eq.(\ref{squ}) remains
valid, is the `symmetrized trace' suggested   by Tseytlin \cite{Ts97}. The
original argument was that in the non-Abelian case the commutators of the field
strength tensors can be expressed via their derivatives, these should be absent
within the constant field approximation in which the NBI action s derived in
the string theory. The trace definition eliminating commutators involves
the symmetrization of all products of the generators obtained in the power
series expansion of the square root. Independently of whether this prescription
corresponds indeed to the string theory result in all orders in $\beta$ (actual
calculations shows that it may be not true in higher
orders~\cite{ReSaTe01,Be01,SeTrTr01}), this is a valuable model to be
considered.

The explicit form of the  SU(2) NBI action with the symmetrized
trace for static $SO(3)$-symmetric magnetic type configurations
was found  in \cite{DyGa00}.  One starts with the definition
\begin{equation} \label{LNBI}
L^{str}_{NBI}=\frac{\beta^2}{4\pi} \ensuremath{\mathop{\rm
Str}\nolimits}\left(1- \sqrt{-\det\Bigl(g_{\mu\nu}+
\frac{1}{\beta}F_{\mu\nu}\Bigr)}\right)= k
\frac{\beta^2}{4\pi}\ensuremath{\mathop{\rm Str}\nolimits}
(1-\mathcal{R}),
\end{equation}
where
\begin{equation}
\mathcal{R}=\sqrt{1+\frac{1}{2\beta^2}F_{\mu\nu}F^{\mu\nu}
-\frac{1}{16\beta^4}(F_{\mu\nu}{\tilde F}_{\mu\nu})^2},
\end{equation}
and $\beta$ of the dimension of length$^{-2}$ is the BI 'critical
field'. Assuming the usual
spherically symmetric t'Hooft-Polyakov ansatz
\be
 A^a_0=0\qquad  A^a_i=\epsilon_{aij}\frac{n^j}{r}(1-\f(r)),\label{aans}
\ee where $n^k=x^k/r$, $r=\sqrt{x^2+y^2+z^2}$ and $\f(r)$ is the
real valued function, the lagrangian (\ref{LNBI}) is
evaluated by expanding the square root in powers of $F$, performing
symmetrization of all products of the gauge generators, evaluating the trace,
and finally summing up the resulting expansion. The result is the following
\cite{DyGa00}:
\begin{equation} \label{Str}
L^{str}_{NBI}=\frac{\beta^2}{4\pi}\left(1-\frac{1+V^2+K^2\mathcal{A}}{\sqrt{1+V^2}}
\right),
\end{equation}
where
\begin{eqnarray}\label{cAdef}
V^2&=&\frac{(1-w^2(r))^2}{\beta^2r^4},\quad
K^2=\frac{w'^2(r)}{\beta^2 r^2},\nonumber\\
\mathcal{A}&=&\sqrt{\frac{1+V^2}{V^2-K^2}}
\ensuremath{\mathop{\mathrm{arctanh}}}\sqrt{\frac{V^2-K^2}{1+V^2}},
\end{eqnarray}
and  dash denotes derivative with respect to $r$.
This form of the lagrangian is appropriate for $V^2>K^2$, otherwise the
function $\ensuremath{\mathop{\mathrm{arctanh}}}$ could be replaced by
$\arctan$. Note that, when the difference
$V^2-K^2$ changes sign, the  function $\mathcal{A}$ remains real valued.

The action with an ordinary trace applied to the square root form
(\ref{squ}) reads
\begin{equation}
L^{tr}_{NBI}=\frac{\beta^2}{4\pi}\left(1-\sqrt{1+V^2+2K^2}\right)\label{Otr}.
\end{equation}
It can be checked that, as $\beta \to \infty$, the standard Yang-Mills
lagrangian (restricted to the monopole ansatz) is recovered in both cases.
But in the strong field region the two expressions differ substantially.

The total action to be used here is the sum $S=S_{NBI}+S_H$,
where the Higgs part
is taken in the usual form \be S_H=
\frac{1}{8\pi}\int\left(D_\mu\phi^a D^\mu\phi^a-\frac{\lambda}{2}
\left(\phi^a\phi^a-v^2\right)\right). \ee Here the dimensionless
gauge coupling constant (in units $\hbar=c=1$) is set to unity,
so we have three parameters: the Higgs expectation value $v$,
Higgs self-interaction constant $\lambda$ and the BI critical
field $\beta$. By additional rescaling, the constant $v$ can be
set to unity. For the Higgs field  the standard hedgehog ansatz is
assumed: \be \phi^a=\frac{H(r)}{r}n^a. \ee

\section{NBI sphalerons inside monopoles}
We start with the square root form of the NBI action (\ref{Otr}).
Performing an integration over spherical angles one obtains the
one-dimensional reduced action, equal to minus the energy
functional:
\be
S = -4\pi
\int dr\, r^2 \left\{
\beta^2({\cal R} - 1) + \frac{1}{2 r^2}
\left(
(H'-\frac{H}{r})^2 + \frac{2}{r^2} H^2\f^2
\right)
+ \frac{\lambda}{4}\left(  \frac{H^2}{r^2} - 1 \right)^2
\right\}, \label{energy}
\ee
where
\be
{\cal R}=\sqrt{1+\frac{2}{\beta^2 r^{4}}(r^2 \f'^{2}+
\frac{1}{\beta^4}(\f^{2}-1)^{2})}.
\ee
Variation of this functional leads to the following equations of
motion
\bea
 r^{2}\f'' &=& \f(\cR H^{2} + \f^{2}-1)+ r^{2}\frac{\cR'}{\cR}\f',
\label{eqf}\\
r^2 H''&=&2H\f^{2}-\lambda H(r^{2}-H^{2})\;. \label{eqh}
\eea
Boundary conditions at infinity for the monopole solutions read
\be
\lim_{r \to \infty} \f(r) = 0, \qquad  \lim_{r \to \infty}
\frac{H(r)}{r}  = 1, \label{infbound}
\ee
while regularity at the origin implies
\be
\f(0) = 1, \qquad H(0) = 0 . \label{orgbound}
\ee
Starting with the values (\ref{orgbound}),
one can construct the following power
series solution converging in some domain around the origin:
\bea
\f &=&1-b r^2+
\frac{\beta b^2 \left(22 b^2 + \beta^2\right)+
d ^2 \left(6 b^2+ \beta^2 \right)^{\frac{3}{2}}}
{10\beta \left(2b^2+ \beta^2 \right)}\,r^4
+ O(r^6), \label{fexp}\\
H &=& d\,r^2-\left(  \frac{1}{10} \lambda d+\frac{2}{5}d\, b \right)r^4
+ O(r^6),\label{hexp}
\eea
where $b$ and $d$ are free
parameters. Solutions which start at the origin in this way reach monopole
asymptotics~(\ref{infbound}) for some discrete values of $b$ and $d$. The equations
reduces to those of the standard YMH-theory as $\beta \to \infty$.
In \cite{GrMoSc99} it
was shown that monopole  solutions to the Eq.(\ref{eqf}-\ref{eqh}),
generalizing the usual t'Hooft--Polyakov monopole,
continue to exist for all finite $\beta$ greater than some limiting value
$\beta_{cr}$ of the order unity. Here we investigate
in more details what happens in the critical region.

The situation resembles that of gravitating monopoles, i.e.  solutions of the
Einstein-Yang-Mills-Higgs equations~\cite{BrFoMa95}. In that case there is a
critical value of the gravitational constant (with other parameters fixed)
after which regular monopoles cease to exist. In the critical region the
gravitational interaction becomes strong and the new solution with the monopole
asymptotics arise, which look like bound states of the monopole with the
Bartnik-McKinnon particles \cite{BrFoMa95}, the latter being solutions of the
Einstein-Yang-Mills equations without Higgs. The bifurcation of the monopole
branch giving rise to excited monopoles occurs in the vicinity of the critical
value of the gravitational constant, the detailed picture depending on the Higgs
mass.

The analogy with gravitating monopoles is based on the fact that in the flat
space NBI theory (without Higgs) there also exist sphaleron type particle-like
solutions~\cite{GaKe99}. In this case the Derrick's obstruction is overpassed
due to violation of the scale invariance by the NBI action.  Therefore, one might
expect the existence of bound states of monopoles with NBI sphalerons inside.
The structure of these solutions can be described as follows.  Near the origin
the Higgs field is almost zero, so the influence of the term $H^2K\cR$ in the
equations of motion is small, and the NBIH system behaves similarly to the NBI
one.  As was argued in \cite{GaKe99}, NBI theories with different $\beta$ are
equivalent up to rescaling,  so, for $\beta$ large enough, the
formation of NBI sphalerons starts close to the origin. Outside their core the
Higgs field becomes significant, and the solution in the far zone is driven to
the monopole configuration.  More precisely, in the region of $r\approx
1/\sqrt{\beta}$, the function $\f(r)$ is similar to that is the sphaleron type
solutions of \cite{GaKe99}: starting  with $\f=1$ it passes through zero and
(possibly) oscillates $N$ times before the solution is driven to an asymptotic
regime. This time, however, the function $\f$ tends not to one of its vacuum
values $\f=\pm 1$, but is captured to the monopole asymptotic $\f=0$. The Higgs
field   $H(r)$ behaves  qualitatively in the same way as in the ground state
monopole solutions.  These considerations can be converted into the rigorous
proof of existence, like in the case of NBI sphalerons~\cite{GaKe99}.

Therefore, for large enough $\beta$, one expects to find the solutions which
can be thought of as bound states of the t'Hooft-Polyakov monopole (slightly
distorted by the Born-Infeld nonlinearity)  and a very small (of size
$\approx\beta^{\frac 12}$)  sphaleron sitting in the monopole core.
Numerically one finds the whole family of excited monopoles which corresponds
to the family of solutions discovered in \cite{GaKe99}. They are labeled by the
number $N$ of nodes of the  function $\f$, the $N=0$ solution being the ground
sate NBI monopole.

The second parameter which enters the lagrangian (\ref{energy}) is the
self-coupling parameter $\lambda$ of the Higgs field. One can see from the
equation of motions that the respective terms are small in the deep core,
where the excited solutions  are starting to form.  In the
region of near critical $\beta$, all happens in the close vicinity of the origin,
so solutions are little sensitive to the values of $\lambda$.
Numerical calculations show that the parameters $b$ and $d$ of the excited
solutions remain practically unchanged when $\lambda$ is doubled.
This is not true for the ground state $N=0$ monopole, which mostly inherits
properties of the Yang-Mills-Higgs solution, but one observes that the role
of the $\lambda$-depending terms also decreases while approaching the critical
region. In particular, the value of  $\beta_{cr}$  is
independent of $\lambda$.

\begin{figure}
\includegraphics[width=14cm,height=11cm]{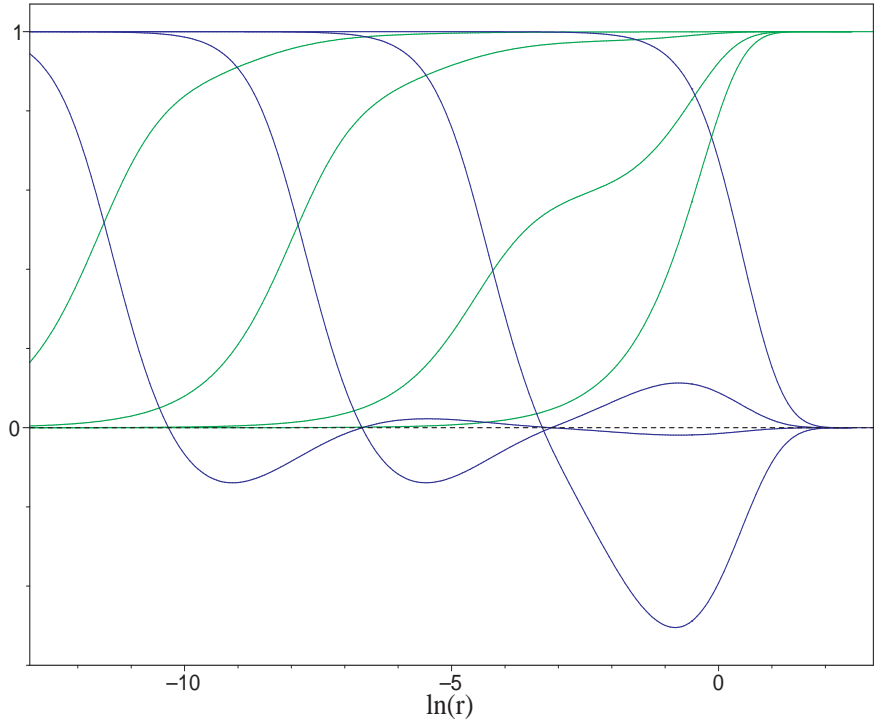}
\caption{Ground state monopole and three first excited states for
$\beta=300$. Blue line --- gauge function $w$, green line ---
Higgs field $H/r$. } \label{fig1}
\end{figure}

Numerical solutions were constructed starting with the regular
initial conditions in the origin (\ref{orgbound}) in terms of the
logarithmic variable $t=\ln(r)$ and applying the shooting
strategy to find the discrete values of parameters $b$ and $d$
which ensure that the monopole asymptotic conditions
(\ref{fexp}-\ref{hexp}) are fulfilled after several oscillations
of $\f$.  The initial guess for the values of  $b$  is  provided
by the glueball values found in~\cite{GaKe99}. Dependence of $b$
on $N$ and $\beta$ is shown on the Tab.~2. The other parameter,
$d$, turns out to be weakly sensitive to the value of $\beta$ and
tends to a constant, as $\beta$ approaches infinity (Tab.~3). The
sample  solutions for $N=1,2,3$  are shown on Fig.~\ref{fig1}
together with the ground state monopole for $\lambda=1$ and
$\beta=30$.

\section{Born-Infeld collapse}

\begin{figure}
\includegraphics[width=14cm,height=11cm]{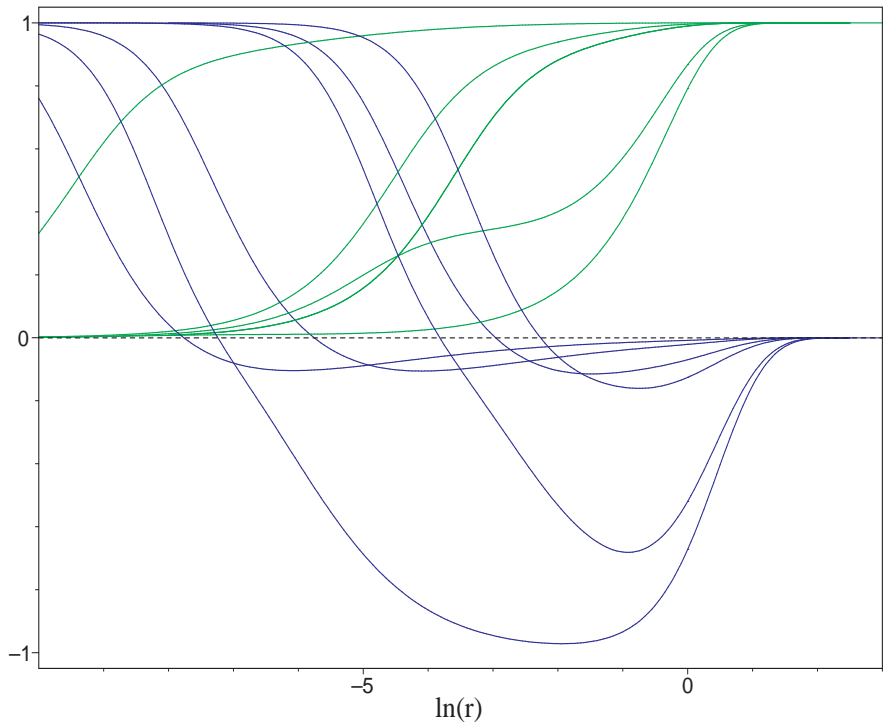}
\caption{The $N=1$ monopole-sphaleron bound state for $\beta=1,
1.05, 1.5, 10, 10^3, 10^6$. The Blue line --- $w$, the green line
--- $H/r$. Solutions  can be distinguished by their values in the
region $\ln r\approx -0.5$:   values of $w$ grow up with
increasing $\beta$, while values of the Higgs field  go down.}
\label{fig2}
\end{figure}

This simple picture holds if the size of sphalerons inside the
monopoles is small as compared with the size of the unexcited
monopole. However, with decreasing $\beta$, the ratio of the
sphaleron radius to that of the monopole increases. At the same
time, both radii rapidly fall down. Numerically it is observed
that for  large enough $\beta$ the parameter $b$ first goes down
with decreasing $\beta$), causing the expansion of the NBI
glueball (the region of $\f$-oscillations).  On the other hand,
the monopole radius stabilizes for sufficiently small $\beta$
(see \cite{GrMoSc99}), so both sizes become comparable for certain
$\beta$.  With $\beta$ further decreasing, the YM parameter $b$
in (\ref{fexp}-\ref{hexp}) starts to increase, as well as the
Higgs parameter $d$. The whole solution then acquires the
following structure: the function $\f$, starting from the value
$\f=1$, rapidly falls down, performs several oscillations of small
amplitude around zero, and  approaches zero asymptotically. The
Higgs variable $H/r$ does not have any peculiarities: starting
from initial value it monotonically grows up to the vacuum
expectation value. For smaller $\beta$, the both parameters $b$
and $d$ begin to grow faster and faster and for the critical
value  $\beta_{cr}$ they tends to infinity.  In terms of the
logarithmic radial variable $t=\ln r$ one clearly sees that
shapes of $\f$ and $H/r$ remain almost the same while $\beta$
approaches the critical value, but the whole picture moves  to
negative $t$ (Tab.~1).  This means that the region of the
localization of essentially nonabelian structure becomes smaller
and smaller, while in the outer region the solution looks like an
embedded Abelian monopole
\begin{equation}
\f= 0 , \qquad \frac H r = 1. \label{absol}
\end{equation}
The figure \ref{fig2} illustrates this behavior for the lowest
excited solution $N=1$. The functions $w$ and $H$ are plotted for
various $\beta$, from very large, up to ones slightly exceeding
$\beta_{cr}$. It is seen that the YM function starts to fall down
for smaller and smaller values of the radial coordinate when
criticality is approached. Similar picture hold for excited
solutions with any number of nodes. This collapse process is
observed also for the ground state monopole $N=0$ near the
critical value $\beta_{cr_0}$, approaching which the solution
shrinks to the abelian counterpart. The numerical experiments
indicate that the critical values of $\beta$ are independent of
$\lambda$. For $N\ge 2$, the value $\beta_{cr_0}$ is practically
independent  on $N$ and is equal to $\beta_{{cr}_2}\approx 1.03$.
For $N=1$ the collapse occurs at $\beta_{{cr}_1}\approx 0.92$
while for the ground state monopole $\beta_{{cr}_0}\approx 0.52$.

Physical reason for the Born-Infeld collapse can be understood as follows.
At high density, the stress-energy tensor of the NBI field develops
large negative pressure. From the lagrangian
(\ref{Otr}) one can derive the following relation for the sum of
principal pressures $p_r=T_{rr},\, p_\phi=T_{\phi\phi}$:
\be
2p_{\phi}+p_{r}=f(\rho)=-\frac{\rho(\rho-2\beta^2)}{\rho+2\beta^2},
\ee
where $\rho=T_{tt}$ is the energy density. This quantity becomes
negative for $\rho>2\beta^2$. Numerical data for the radial and
tangential pressures corresponding to the ground state monopole
are given in Figs. \ref{fig4},\ref{fig5}  for the pure gauge component
and together with  Higgs. When $\beta$ goes to the critical
value, the tangential pressure exhibits a small positive knob
near the monopole boundary and then becomes negative in the core
region. With decreasing $\beta$ it falls down rapidly tending to
minus infinity at the threshold. The radial pressure near the
threshold is always negative and also tend to minus infinity.

\begin{figure}
\includegraphics[width=14cm,height=11cm]{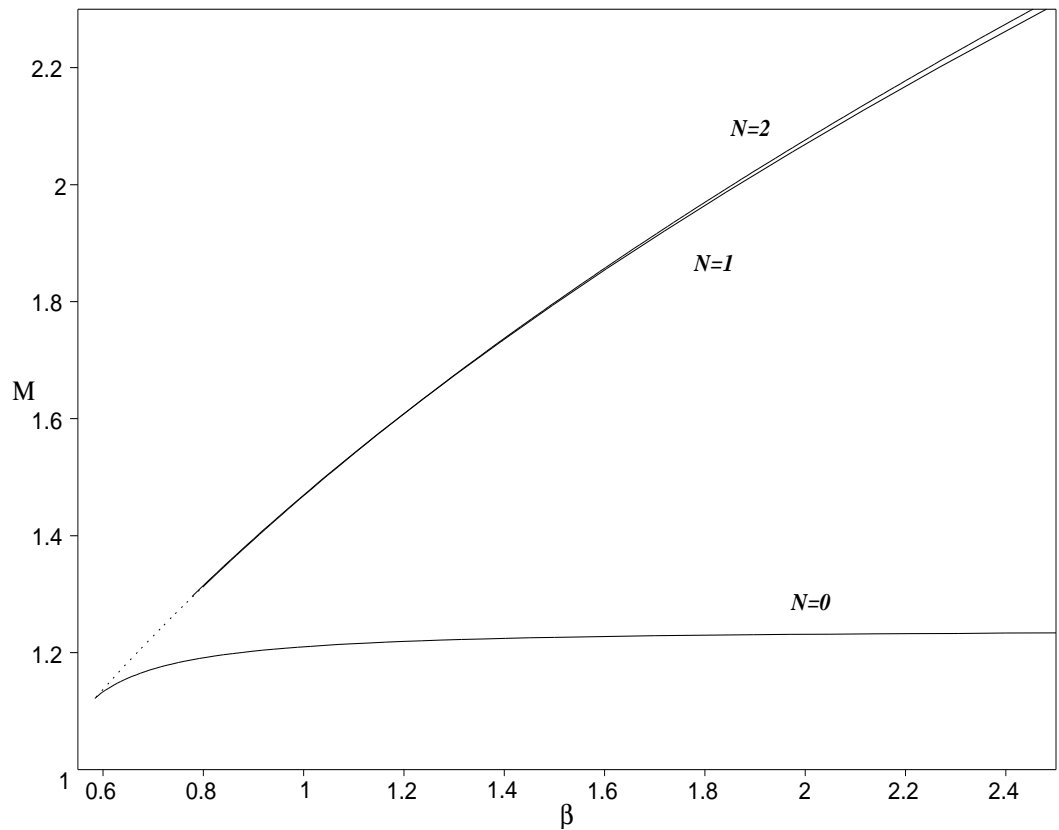}
\caption{The masses of the ground state monopole $(N=0)$ and $N=1, 2$
excited monopoles. The dotted line is the mass of embedded
abelian solution (\ref{abmass}).  The $N=2$ mass curve, in the
region where this solution exists,
is indistinguishable  from mass of the embedded abelian
monopole.}
\label{fig3}
\end{figure}

\begin{figure}
\includegraphics[width=14cm,height=11cm]{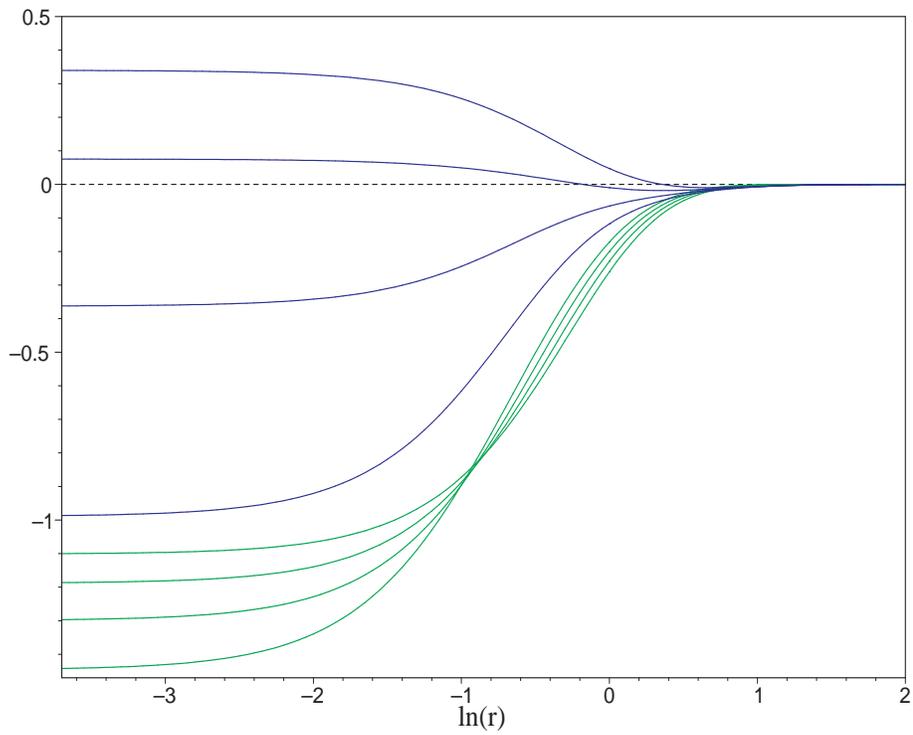}
\caption{ The radial pressure  $p_{r}$ for the ground state
BI-monopole for  $\beta = 0.9, 1.0, 1.2, 2.0 $.  The blue line
--- the pressure of the NBI field, the green line --- the
pressure of the Higgs field. The left asymptotic values go down
with decreasing $\beta$ for both fields. }\label{fig4}
\end{figure}

\begin{figure}
\includegraphics[width=14cm,height=11cm]{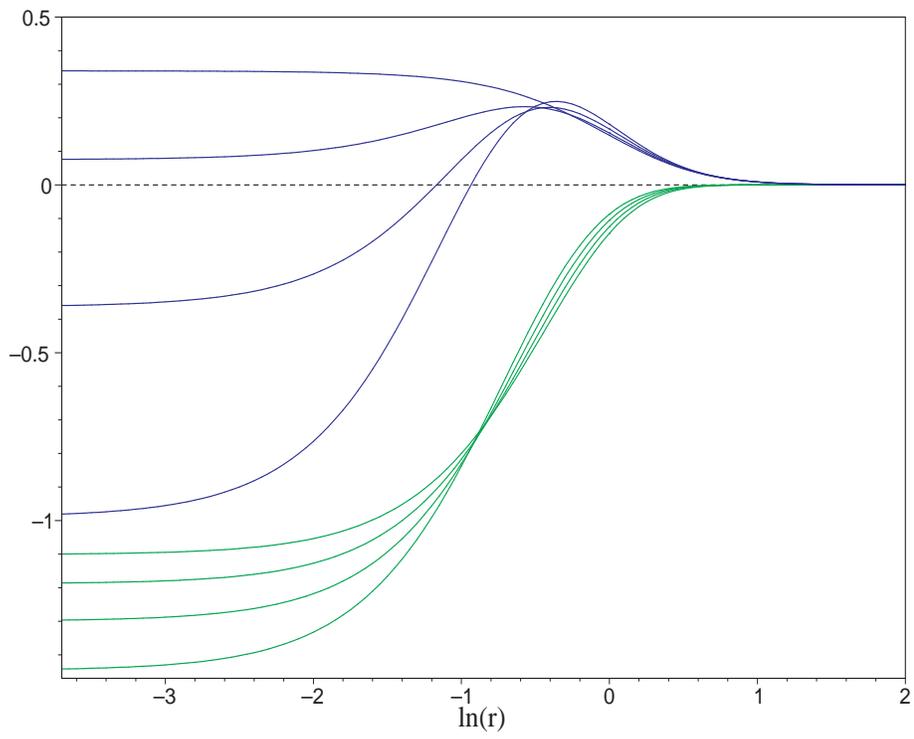}
\label{fig5} \caption{The tangential pressures $p_{\phi}$ behave
similarly to those on previous picture }
\end{figure}

The masses of solutions in the critical region rapidly converge to the
energy of the embedded Abelian pointlike monopole (recall that
in the Born-Infeld theory the singular pointlike solution \ref{absol}
has a finite energy):
\begin{equation}
 E_{lim} =\int dr\,r^2 2\beta^2 (\cR-1)=\sqrt{\beta}
\int dx\,x^2(\sqrt{1+\frac{2}{x^4}}-1)=
1.23604978\sqrt{\beta}.\label{abmass}
\end{equation}
This value serves as the upper bound for the mass as a function of
$\beta $ and $N$.
Moreover, this value describes pretty well the masses of all excited solutions
starting from $N=1$: for all $\beta$ the discrepancy does
not exceed 3\% and it decreases fast near the threshold.
This can be seen on Figure \ref{fig3}, showing the masses of $N=0,1,2$
solutions as a function of $\beta$. The bifurcation pattern of the main
branch with excited branches is similar to that
of gravitating monopoles, but now it occurs with a decreasing parameter.

\section{Symmetrized trace model}
Within the symmetrized trace model
the equations of motion read:
\[
\frac{d}{dr}\left\{\frac{w'}{2(V^2-
K^2)}\left(\frac{K^2\sqrt{1+V^2}}{1+K^2}-
\frac{(2V^2-K^2)\cA}{\sqrt{V^2-K^2}}\right)\right\} =\]
\be
\label{eqw}
 \frac{Vw(K^2\cA -V^2)}{(V^2-K^2)\sqrt{1+V^2}}+\frac{w\,H^2}{r^2}
\ee
\be
r^2 \,H''=H(H^2-r^2)\lambda+2Hw^2,
\ee
where $V$ and $K$ are defined in (\ref{cAdef}).

\begin{figure}
\includegraphics[width=14cm,height=11cm]{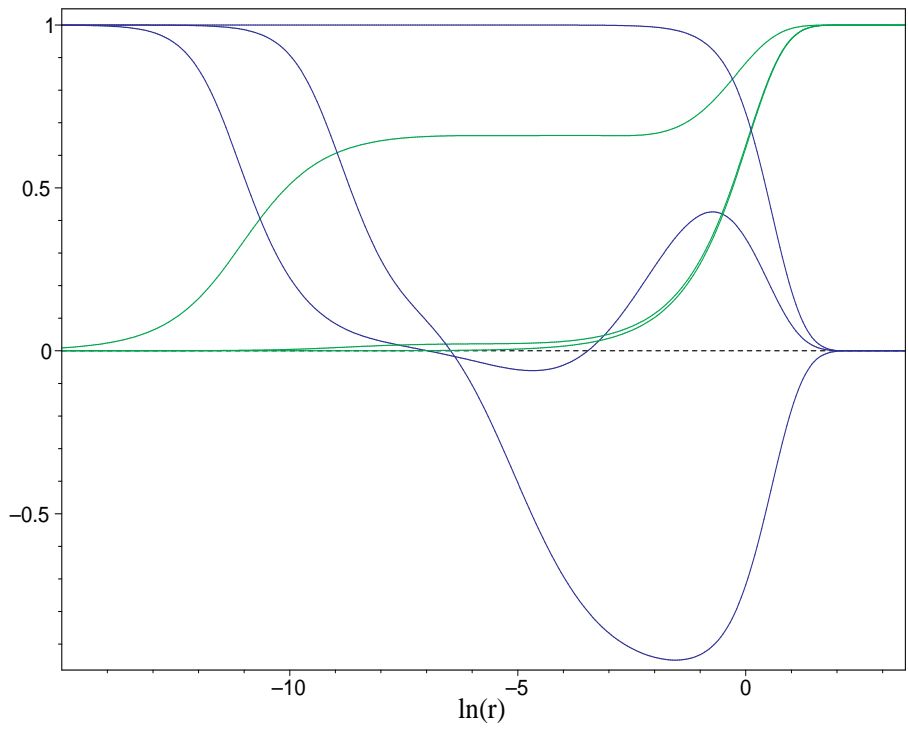}
\label{fig6} \caption{The monopole-sphaleron bound states in the
symmetrized trace model for $\beta=10^6 $. Blue line --- gauge
function $w$, green line --- Higgs field $H/r$. }
\end{figure}

The pure NBI model with the symmetrized trace also possesses the sphaleronic
type excitations which are qualitatively the same as in the ordinary trace
model \cite{DyGa00}. In particular, the same scaling behavior follows
immediately from the action (\ref{Str}).
Hence the above arguments about the existence of monopole-sphaleron
bound states apply directly, and one can  expect to find sphaleronic
excitations at least for large $\beta$.
Numerical experiments   indicate that for large
values of $\beta$ there are indeed excited monopole solutions
analogous to those considered above (Fig.~\ref{fig6}).
But, since the structure of the essentially nonabelian region is not the same
in both models, there are some differences in the behavior of the solution
with varying $\beta$. One observes that, in the symmetrized
trace model, the sphaleronic solutions for a given node number $N$ are
smaller than those in the ordinary trace model.
Also, the amplitude of $\f$-oscillations is relatively smaller.
All this is reflected in the relatively large values of
the parameters $b$ and $d$ which now grow much faster with
increasing $\beta$ and $N$.

As a consequence, the collapse, which takes place in this model also,
occurs at  larger values of $\beta$ for any positive $N$.
Moreover, for any given $\beta$, there exist excited solutions
only up to some finite $N$, the boundary value of $N$ being
dependent on $\beta$.

\section{Discussion}
Our purpose was to describe an astonishing similarity between gravitating
monopoles and those in the flat space non-Abelian Born-Infeld theory.
Both theories contain ground state monopoles as well as excited monopoles
which can be regarded as bound states of monopoles and sphalerons. In both
cases there is a threshold in the parameter space, which marks the end of
the monopole sequence. In the gravitating case this happens since the
monopole radius becomes smaller that its gravitational radius, so it is the
gravitational collapse phenomenon which stands behind this picture.
Physically, gravitational collapse can be attributed to development of strong
attractive forces when the event horizon is approached.
Surprisingly enough, we observed an analogous behavior in
in the flat space NBI theory. The size of monopoles (including their excited
states) become smaller and smaller when a certain threshold value of the
BI field strength parameter is approached, and no extended monopoles exist
beyond the threshold. In this region the large negative pressure (tension)
is developed inside the regular NBI monopoles, which cause them to shrink
to a singular pointlike (Abelian) monopole.  Large negative pressure
developed inside the high density ball of the NBI matter apparently should
create an instability against the contraction to the pointlike structure.
The dynamical picture of the collapse phenomenon in the NBI theory is currently
under investigation.

It is worth noting, that if one couples the NBI model to gravity,
a continuous interpolation between the NBI sphalerons and
Bartnik-McKinnon particles is observed \cite{DyGa'00}. It is
expected that parameters of monopole-sphaleron bound states will
also exhibit the analogous transition.


\newpage
\begin{table}
\centering
\begin{tabular}{|c|c|c|c|c|c|}
\hline $\beta$   &   $N=0$          &   $r_w, N=1$  & $r_h, N=1$        &   $r_w, N=2$  &   $r_h, N=2$ \\
\hline $10^6$    & 1.399        & $.7175\cdot10^{-3}$   & .5082         & $.9212\cdot 10^{-4}$  & .4970        \\
\hline $10^4$    & 1.399        & $.7152\cdot10^{-2}$   & .4512         & $.8348\cdot 10^{-3}$  & .2970        \\
\hline 100   & 1.399        & .05451        & .01876            & $.1529\cdot 10^{-2}$  & $.4325\cdot 10^{-3}$ \\
\hline 10    & 1.397        & .1036         & .02614        & $.2335\cdot 10^{-2}$  & $.6372\cdot 10^{-3}$ \\
\hline 5     & 1.391        & .1126         & .02692        & $.2081\cdot 10^{-2}$  & $.5500\cdot 10^{-3}$ \\
\hline 2     & 1.349        & .08349        & .01790        & $.6481\cdot 10^{-3}$  & $.1542\cdot 10^{-3}$ \\
\hline 1.5   & 1.303        & .05188        & .01058        & $.1690\cdot 10^{-3}$  & $.3766\cdot 10^{-4}$ \\
\hline 1.2   & 1.235        & .01930        & $.3783\cdot 10^{-2}$  & $.1102\cdot 10^{-4}$  & $.2285\cdot 10^{-5}$ \\
\hline 1.15  & 1.215        & .013325       & $.2595\cdot 10^{-2}$  & $.3863\cdot 10^{-5}$  & $.7881\cdot 10^{-6}$ \\
\hline 1.1   & 1.191        & $.7755\cdot 10^{-2}$  & $.1502\cdot 10^{-2}$  & $.8085\cdot 10^{-6}$  & $.1620\cdot 10^{-6}$ \\
\hline 1.    &  1.124       & $.4249\cdot 10^{-3}$  & $.8130\cdot 10^{-4}$  & --- & --- \\
\hline .97   & 1.096            & $.7122\cdot 10^{-5}$  & $.1350\cdot 10^{-5}$  & --- & --- \\
\hline .8    & .7936        & ---           & --- & --- & ---   \\
\hline .7    & .4267        & ---           & --- & --- & ---   \\
\hline .6    & .1060        & ---           & --- & --- & ---   \\
\hline .55   & .02596       & ---           & --- & --- & ---   \\
\hline .526  & $.2739\cdot10^{-2}$  & ---           & --- & --- & ---   \\
\hline
\end{tabular}

\caption{This table illustrates the behavior of effective sizes of the monopole and
the monopole-sphaleron bound states $N=1,2$ with decreasing $\beta$. The sizes are defined
as follows:
for N=0 it is the distance where the gauge function falls down by half, i.e. $\f(r)=0.5$;
for N=1,2  the 'sphaleron radius' $r_w$ is the first node of $w$, while the 'monopole radius'
$r_h$ is the distance where $H(r)/r=.5$}
\end{table}


\begin{table}
\centering
\begin{tabular}{|c|c|c|c|}
\hline  $\beta$   & $N=0$    & $N=1$ & $N=2$ \\
\hline  $10^6$    & .4496 & $.18011\cdot10^{8}$ &  $.1258\cdot10^{10}$ \\
\hline  $10^4$    & .4496 & $.1816 \cdot 10^6$ &  $.1533\cdot10^{8}$ \\
\hline  100       & .4496 & 3448. &  $.4611\cdot10^{7}$ \\
\hline  10        & .4519 & 1089. &  $.2141\cdot10^{7}$ \\
\hline  5         & .4592 & 1027. &  $.2946\cdot10^{7}$ \\
\hline  2         & .5198 & 2598. &  $.4073\cdot10^{8}$ \\
\hline  1.5       & .5978 & 8100. &  $.7163\cdot10^{9}$ \\
\hline  1.2       & .7496 & $.6868\cdot 10^5 $ &  $.2046\cdot10^{12}$ \\
\hline  1.15      & .8007 & $.1524\cdot 10^6 $ &  $.1739\cdot10^{13}$ \\
\hline  1.1       & .8677 & $.4673\cdot 10^6 $ &  $.4169\cdot10^{14}$ \\
\hline  1         & 1.086 & $.1699\cdot 10^9 $ &  --- \\
\hline  .97       & 1.189 & $.6261\cdot 10^{12}$ &  ---  \\
\hline  .8       & 2.951 & --- & --- \\
\hline  .7       & 9.954 & --- & --- \\
\hline  .60      & 142.1 & --- & --- \\
\hline  .55      & 2314. & --- & --- \\
\hline  .526     & $.2070\cdot 10^6$ & --- & --- \\
\hline
\end{tabular}

\caption{The numerical values of the parameter $b$ of the expansion
of  $\f$   near the
origin for various   $\beta$ }
\end{table}

\begin{table}
\centering
\begin{tabular}{|c|c|c|c|}
\hline  $\beta$   & $N=0$    & $N=1$ & $N=2$ \\
\hline  $10^6$    &  1.068 &  35.69 &  792.3 \\
\hline  $10^4$    &  1.068 &  36.20 &  896.1 \\
\hline  100       &  1.068 &  36.70 &  1483. \\
\hline  10        &  1.069 &  23.88 &  1007. \\
\hline  5         &  1.072 &  23.15 &  1171. \\
\hline  2         &  1.096 &  35.41 &  4229. \\
\hline  1.5       &  1.124 &  60.74 &  $.1744\cdot 10^5$ \\
\hline  1.2       &  1.173 &  171.8 &  $.2898\cdot 10^6$ \\
\hline  1.15      &  1.188 &  253.0 &  $.8420\cdot 10^6$ \\
\hline  1.1       &  1.207 &  439.3 &  $.4105\cdot10^7$ \\
\hline  1.        &  1.264 &  8215. &  --- \\
\hline  .97       &  1.289 &  $.4962\cdot 10^6$ &  --- \\
\hline  .8       & 1.638 &  --- & --- \\
\hline  .7       & 2.498 &  --- & --- \\
\hline  .6       & 7.838 &  --- & --- \\
\hline  .55      & 29.63 &  --- & --- \\
\hline  .526     & 273.3 &  --- & --- \\
\hline
\end{tabular}

\caption{The numerical values of the parameter $d$ of the Higgs field expansion
near the origin for various  $\beta$}
\end{table}

\end{document}